\documentclass{aastex}
\usepackage{emulateapj5}

\slugcomment{}

\shorttitle{Spatial study with the VLT of a new resolved edge-on disk}
\shortauthors{Grosso et al.}

\begin{document}


\title{Spatial study with the VLT\\
of a new resolved edge-on circumstellar dust disk\\
discovered at the periphery of the $\rho$ Ophiuchi dark cloud\altaffilmark{1}}


\author{N. Grosso\altaffilmark{2}, J. Alves\altaffilmark{3}, K. Wood\altaffilmark{4}, R. Neuh{\"a}user\altaffilmark{2}, T. Montmerle\altaffilmark{5}, and J.~E. Bjorkman\altaffilmark{6}}


\altaffiltext{1}{Based on observations carried out at the European Southern 
Observatory, La Silla, Chile, under project 67.C-0325(A), and Paranal, Chile, 
under service mode project 267.C-5699(A).}
\altaffiltext{2}{Max-Planck-Institut f{\"u}r extraterrestrische Physik,
      	         P.O. Box 1312, D-85741 Garching bei M{\"u}nchen, Germany;\\
		 ngrosso@xray.mpe.mpg.de}
\altaffiltext{3}{European Southern Observatory,
	         Karl-Schwarzschild-Str. 2,
       	         D-85748 Garching bei M{\"u}nchen, Germany}
\altaffiltext{4}{School of Physics \& Astronomy, University of St Andrews, 
                 St Andrews, KY16 9SS, Scotland}
\altaffiltext{5}{Service d'Astrophysique, 
     	         CEA Saclay, 
	         F-91191 Gif-sur-Yvette, France}
\altaffiltext{6}{Ritter Observatory, Department of Physics \& Astronomy, 
                 University of Toledo, OH 43606, USA}


\begin{abstract}
We report the discovery in near-infrared (NIR) with SofI at the 
{\it New Technology Telescope} (NTT) of a resolved circumstellar dust disk 
around a 2MASS source at the periphery of the $\rho$ Ophiuchi dark cloud. 
We present follow-up observations in $J$, $H$, and $K_\mathrm{s}$-band obtained with 
ISAAC at the {\it Very Large Telescope} (VLT), under 0\farcs4-seeing conditions, which unveil 
a dark dust lane oriented East-West between two characteristic northern and southern 
reflection nebulae. 
This new circumstellar dust disk has a radius of 2\farcs15~(300\,AU at 140\,pc), 
and a width of 1\farcs2~(170\,AU at 140\,pc).
Thanks to its location at the periphery of the dense cores,
it suffers small foreground visual extinction ($A_\mathrm{V}= 2.1\pm2.6$\,mag).
Although this disk is seen close to edge-on, the two reflection nebulae display 
very different colors. 
We introduce a new NIR data visualization called ``Pixel NIR Color Mapping'' 
(PICMap for short), which allows to visualize directly the NIR colors of the nebula pixels. 
Thanks to this method we identify a ridge, 0\farcs3 (40\,AU at 140\,pc) 
to the north of the dark lane and parallel to it, which displays a NIR color excess. 
This ridge corresponds to an unusual increase of brightness from $J$ to $K_\mathrm{S}$, 
which is also visible in the NTT observation obtained 130\,days before the VLT one.
We also find that the northern nebula shows $\sim$3\,mag more extinction than 
the southern nebula.
We compute axisymmetric disk models to reproduce the VLT scattered light images 
and the spectral energy distribution (SED) from optical to NIR. 
Our best model, with a disk inclination $i=86\pm1^\circ$, correctly reproduces 
the extension of the southern reflection nebula, but it is not able to reproduce 
either the observed NIR color excess in the northern nebula 
or the extinction difference between the two reflection nebulae.
We discuss the possible origin of the peculiar asymmetrical NIR color properties of this object.
\end{abstract}




\section{Introduction}

Observations of resolved circumstellar disks around young stars are very important 
for our understanding of the formation of solar-type stars and planetary systems.
Usually the large difference of brightness between the central star 
and its circumstellar material does not allow direct imaging of the circumstellar disk.
However when the disk midplane is seen close to edge-on, models predict that 
the direct light of the star is blocked out while a small amount of the light 
is scattered by the grains below and above the disk midplane, producing a typical pattern 
of a dark lane between two reflection nebulae (\citealt{whitney92}). 
The first young stellar object displaying this typical pattern, 
HH\,30\,IRS in the Taurus dark cloud, was imaged by the HST/WFPC2 (\citealt{burrows96}). 
Edge-on disks have since become accessible to ground-based NIR observations thanks to 
adaptive optics or speckle imaging (e.g., observations of \objectname[]{HK\,Tau/c}; \citealt{stapelfeldt98}; 
\citealt{koresko98}).  
Recently new observations of edge-on disks were reported in one of the nearest 
star-forming regions, the $\rho$ Ophiuchi dark cloud (L1688; $d \sim 140$\,pc), both by 
the VLT/ISAAC under 0\farcs35-seeing conditions (\citealt{brandner00}) and the HST/NICMOS 
(\citealt{terebey01}; \citealt{allen02}). 
These disks are all located in the dense cores of this dark cloud and are deeply 
embedded~(Fig.~\ref{chart}). 

During the period April 4--7 2001, five deep NIR pointings 
were obtained with the NTT/SofI in the $\rho$ Ophiuchi 
star-forming region (Grosso et al., in preparation), as follow-up 
of X-ray observations made with Chandra and XMM-Newton. 
\cite*{grosso01} already reported  
the serendipitous discovery of new embedded Herbig-Haro objects in one of these pointings.
We focus here on another pointing at the periphery of 
the $\rho$ Ophiuchi dark cloud (Fig.~\ref{chart}) where we have 
discovered a new resolved edge-on circumstellar dust disk.
We present in $\S$\ref{section_obs} the NTT/SofI data and 
the follow-up observation made with the VLT/ISAAC (Unit Telescope 1; {\it Antu}) 
unveiling the shape of this object.
We measure its photometry in $\S$\ref{section_photometry}, and discuss the foreground visual 
extinction in $\S$\ref{section_extinction}.
We introduce in $\S$\ref{section_color} the Pixel NIR Color Mapping of the two nebulae. 
In $\S$\ref{section_model} axisymmetric models are used to estimate 
the physical parameters of this disk, and to predict NIR colors.
Finally in $\S$\ref{section_discussion} we discuss the possible origin of the 
peculiar NIR color properties of this object.

\begin{figure*}[t]
	\epsscale{2.1}
	\plotone{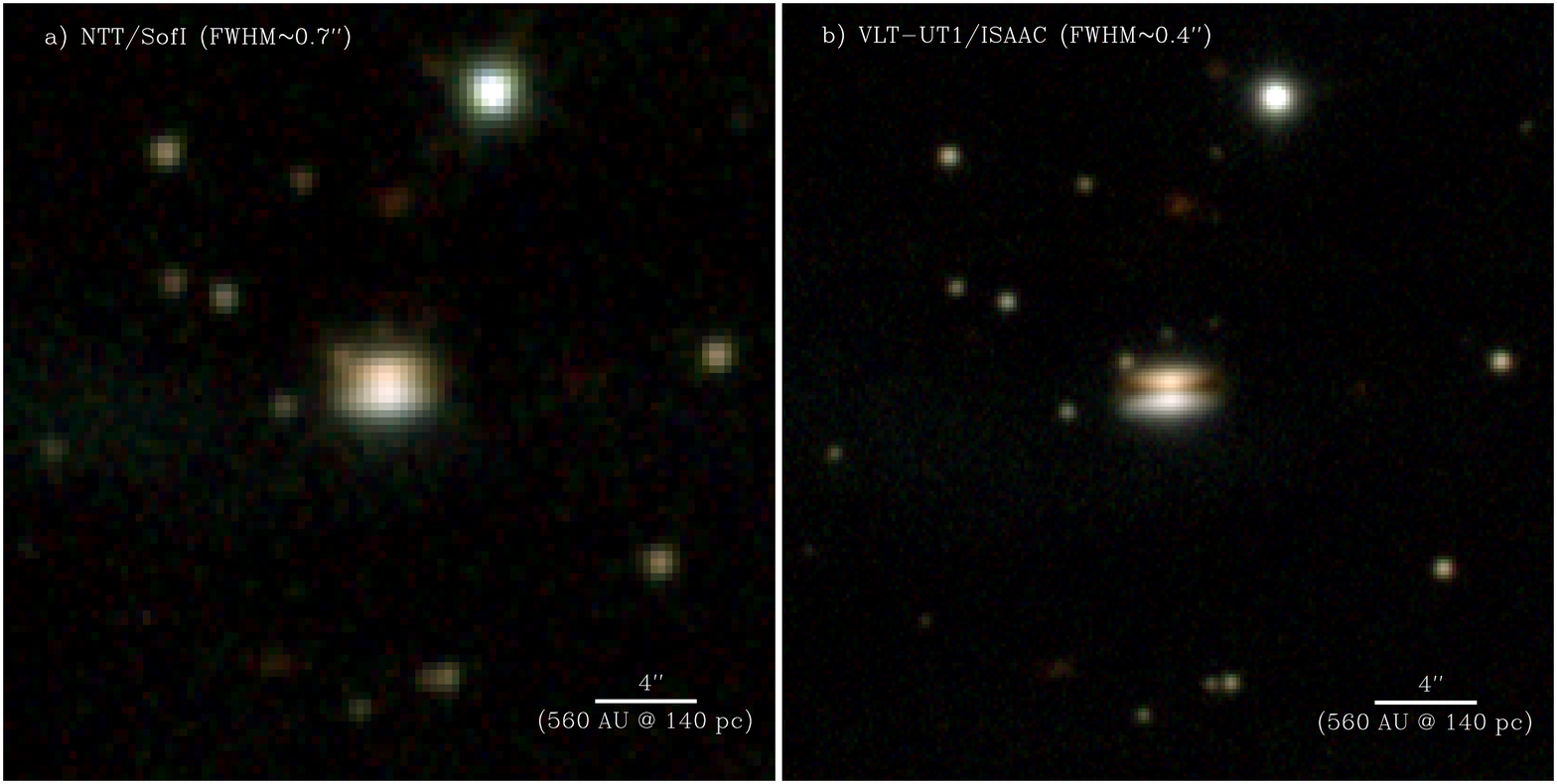}
 \caption{Near-infrared view of 2MASSI\,1628137-243139. 
These color composite images combine  $J$, $H$, and $K_\mathrm{S}$
using respectively blue, green, and red color coding, with logarithmic color stretch.
The white balance is obtained from the brightest background star.
The image size is 30\arcsec$\times$30\arcsec. 
The image orientation is North on the top and East on the left hand side. 
a) observations with the NTT/SofI  
(pixel size = 0\farcs292, seeing FWHM$\sim$0\farcs7).
b) observations with the VLT-UT1/ISAAC  
(pixel size = 0\farcs148, seeing FWHM$\sim$0\farcs4).
}
\label{image}
\notetoeditor{Fig.~\ref{image} must be printed in color on two columns. 
Could you please convert the files to CMYK?}
\end{figure*}

{
\vspace{0.25cm}
   \plotone{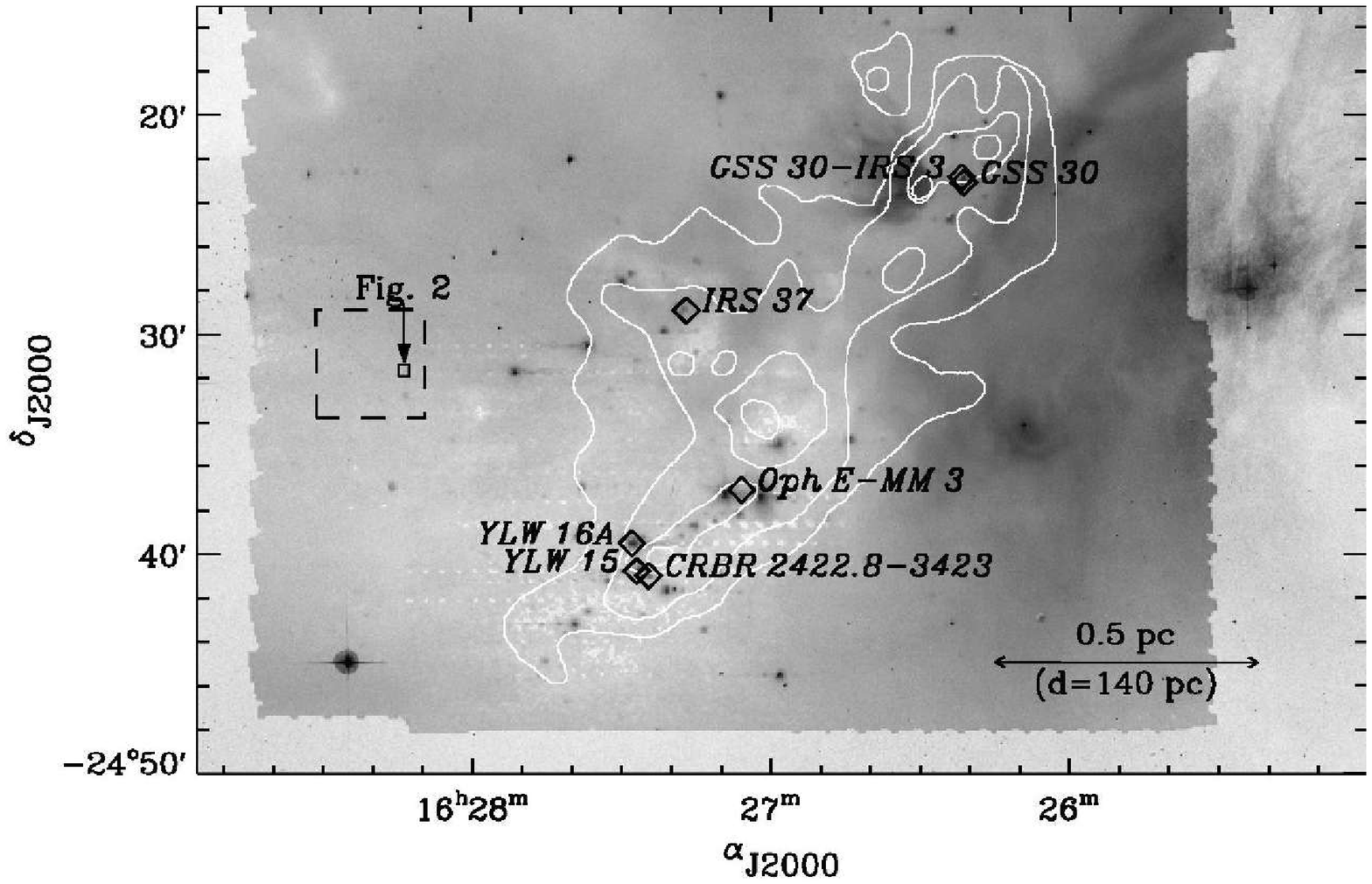}
      \figcaption{Finding chart of the edge-on circumstellar disks in the $\rho$ 
Ophiuchi dark cloud.
The optical background image from the Digitized Sky Survey is merged with 
the ISOCAM mid-infrared map (6.7 and 14.3\,$\mu$m; Abergel et al. 1996). 
The overlaid white contours show the C$^{18}$O column density 
($A_\mathrm{V}$=36, 54, 72, 90\,mag; Wilking \& Lada 1983), 
associated with the L1688 dark cloud.
Black diamonds mark the previously known young stellar objects 
displaying in scattered light two reflection nebulae 
separated by a dark lane (Brandner et al. 2000; Terebey 2001; Allen et al. 2002). 
They are mostly protostars surrounded by remnant infalling envelopes.
The dashed box displays the 5\arcmin$\times$5\arcmin~field of view 
of the NTT/SofI pointing at the periphery of the dark cloud.
The small square box defines the area of Fig.~\ref{image}.
}
\label{chart}
}

\section{Observations}
\label{section_obs}

	\subsection{The NTT/SofI observation}

This observation was made during the night of April 7 at airmass 1.0, 
with the large SofI field ($5^\prime\times5^\prime$; pixel size = 0\farcs292) 
to survey the largest area. We took 60\,s-exposure frames using 
10\,s detector sub-integration time to avoid detector non-linearity and saturation. 
Between each frames the telescope was moved inside a 20\arcsec-width box centered 
on the target position, according to a random pattern of offsets automatically determined 
by the imaging mode called auto jitter.
The net exposure times are 24\,min in $J$, $H$, and $K_\mathrm{S}$, 
leading to a $K_\mathrm{S}$$\sim$19\,mag completeness magnitude for point sources.
The data reduction was performed using the ESO's {\tt eclipse} package (version 3.8.1) 
which provides the {\tt jitter} command.
The best seeing conditions of this NTT/SofI pointing were achieved during 
the $K_\mathrm{S}$ exposure with Full Width at Half Maximum (FWHM) $\sim$0\farcs7.
The astrometric registration of the reduced images was made using 
the position of a $K_\mathrm{S}$=14.1\,mag star (2MASSI\,1628133-243127; 
\citealt{cutri00}) located only 12\arcsec~to the north of our object, 
extracted using SExtractor (\citealt{bertin96}).
$J$ and $H$ images were then resampled to match the $K_\mathrm{S}$ image 
using the {\tt warping} command of {\tt eclipse} with the hyperbolic tangent 
interpolation kernel.

Figure~\ref{image}.a shows the resulting color composite image centered on 2MASSI\,1628137-243139
which went unnoticed until now.
Compared to the unresolved field stars, this source appears extended.
One can barely see two characteristic reflection nebulae, and a marginally 
resolved dark dust lane oriented East-West in front of the star. 
To confirm this detection and to measure the physical parameters of 
this disk, we need a deeper and higher angular resolution image. 
This can easily be achieved with the VLT-UT1/ISAAC, by combining better 
seeing conditions with better sensitivity and image sampling. 
We obtained for this goal one hour of the ESO Director's Discretionary Time.

	\subsection{The VLT-UT1/ISAAC observation}

The observation was carried out in service mode, 
triggered by very good seeing conditions (FWHM$\sim$0\farcs4) on August 15, 2001.
The auto jitter imaging mode was also used with 80\,s, 60\,s, 60\,s-exposure frames 
and 40\,s, 15\,s, 10\,s detector sub-integration time for $J$, $H$, and $K_\mathrm{S}$, 
respectively. 
Net exposure times were 10\,min in $K_\mathrm{S}$ and $H$, and 10.7\,min in $J$, 
with airmass ranges from 1.05 to 1.11.
The data reduction was performed using the version 4.1.0 of {\tt eclipse}.
We used the {\tt is\_ghost} command to remove the electrical ghosts induced by a bright star 
in the field of view.
The photometric calibration is based on the observation of the IR standard star 
9181/S243-E (\citealt{persson98}) provided by the service mode, 
and taking into account average IR atmospheric extinction on Paranal 
(0.11, 0.06, and 0.07\,mag\,airmass$^{-1}$ for $J$, $H$, and $K_\mathrm{S}$, respectively). 
We used the same method as described above for our NTT data to produce 
the resulting color composite image presented in Fig.~\ref{image}.b 
(see also the ESO Press Release, 
Photo 12b/02\footnote{\url{http://www.eso.org/outreach/press-rel/pr-2002/pr-09-02.html}},
which presents a larger color composite image corresponding to the field of view 
shown in Fig.~\ref{extinction_map}).  

Thanks to very good seeing conditions and smaller detector pixel size, 
the two reflection nebulae are unveiled, 
and the dark dust lane is now clearly resolved in all three filters.
The shape of this object seen nearly edge-on and appearing on a starry background 
has led us to nickname it the {\it Flying Saucer} (hereafter the FS). 
Two sources are visible within 2\arcsec~(280\,AU at 140\,pc) of this disk both in 
the NTT and the VLT images. The first source located at the North-East is the brightest 
with $K_\mathrm{S}\sim17.5$\,mag.
Stars with similar colors are visible in Fig.~\ref{image}.b; this source is likely
a background star. The second source, thanks to its location close to the South-North 
disk axis, could be a jet H$_2$ knot candidate, although no counter knot is visible. 
Several objects with similar colors and brightnesses are also spread over the field 
of view, which allows us to interpret this second source also as a background source.

\section{Photometry of the FS}
\label{section_photometry}

\begin{figure*}[t]
     \plotone{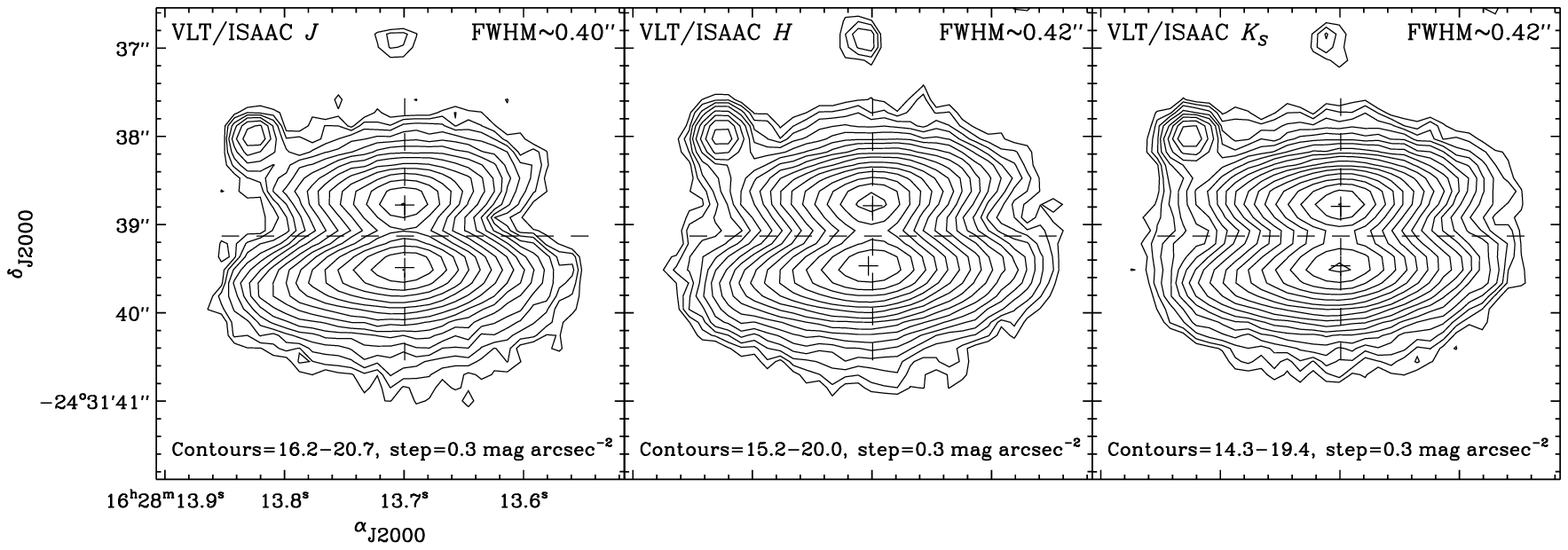}
      \caption{Surface brightness maps of the Flying Saucer from the VLT-UT1/ISAAC observations.               
We give the calibrated contour intensity range from the peak value to the 
first contour, with the contour step.
The crosses mark the peak positions of the reflection nebulae in each filter.
The symmetry axes (dashed lines) are defined from the peak positions in 
the $K_\mathrm{S}$-band image. Their lengths give 
the width and the height of the disk.}
         \label{brightness}
\notetoeditor{Fig.~\ref{brightness} must be printed on two columns.}
   \end{figure*}

Figure~\ref{brightness} presents the surface brightness maps of the VLT data 
in the $J$, $H$, and $K_\mathrm{S}$-band, respectively, with contours 
obtained without any smoothing with the Interactive Data Language ({\tt IDL}).
We define the dark lane axis as the bisector of the line segment defined 
by the nebula peaks in the $K_\mathrm{S}$ image. 
The orientation of the dark lane is strictly East-West, and separates 
the northern and southern reflection nebulae.
The southern nebula is slightly asymmetric, with a larger extension 
towards the East. In $J$ the southern nebula is more extended than its northern counterpart,
and the southern peak is 1.7 times brighter than the northern one. 
This implies that we are not seeing the FS exactly edge-on. 
Indeed the scattered light models of accretion disks versus inclination show that 
when one looks above the disk midplane, the brightness of the upper reflection 
nebula increases compared to the brightness of the lower reflection nebula 
(see \citealt{whitney92}). We are therefore seeing the FS slightly 
below its disk midplane, corresponding to the southern nebula.
The difference between the northern and the southern reflection nebula decreases 
with increasing wavelength, leading to a more symmetric reflection nebulae in $K_\mathrm{S}$. 
In this filter the northern nebula is still slightly less extended, 
but its peak brightness is 1.3 times brighter than the southern nebula.
This produces the reddish aspect of the northern nebula visible in Fig.~\ref{image} 
both in the NTT and the VLT images.
This resolved circumstellar disk has a radius of 2.15\arcsec~(300\,AU at 140\,pc), 
as measured from the largest extension of the southern nebula 
(contour 20.4\,mag\,arcsec$^{-2}$ in $J$), and a width of 1.2\arcsec~(170\,AU at 140\,pc).
The shape of the FS, due to its low disk inclination, is reminiscent 
of HK\,Tau/c~(\citealt{stapelfeldt98}; \citealt{koresko98}), but three times larger 
and without strong asymmetry in the scattered light intensity. 

\begin{table*}[!ht]
\tabletypesize{\scriptsize}
\caption{Optical/NIR integrated photometry of the Flying Saucer.}
\label{photometry}
\begin{tabular}{cccccccc}
\hline
\hline
\noalign{\smallskip}
$B_\mathrm{J}^{\tablenotemark{a}}$ & $B^{\tablenotemark{b}}$ & $R_\mathrm{53F}^{\tablenotemark{a}}$ & $R^{\tablenotemark{c}}$ & $I^{\tablenotemark{d}}$ & $J$ & $H$ & $K_\mathrm{S}$\\
\noalign{\smallskip}
\hline
\noalign{\smallskip}
20.4$\pm$0.3    & 20.5$\pm$0.3    &  18.6$\pm$0.3      & 18.4$\pm$0.3     & 17.16$\pm$0.14     & 15.93$\pm$0.06     & 14.78$\pm$0.05     & 13.98$\pm$0.03\\
\noalign{\smallskip}
\hline
\end{tabular}
\tablenotetext{\mathrm{a}}{SuperCOSMOS Sky Survey (\citealt{hambly01}).}
\tablenotetext{\mathrm{b}}{Derived from $B_\mathrm{J}$ (Kodak IIIa-J plate + GG395 filter) 
using the color correction of \cite*{blair82} assuming the color index $V-R$ corresponding 
to $T_{\rm eff}=3500$\,K in the MK temperature scale (\citealt{cox00}). 
This leads to $V=19.9$ (marked with a white dot in Fig.~\ref{sed}).}
\tablenotetext{\mathrm{c}}{Derived from $R_\mathrm{53F}$ (Kodak IIIa-F plate + RG630 filter) 
using the color corrections of \cite*{bessell86} with $I$ given by DENIS (\citealt{epchtein99}).}
\tablenotetext{\mathrm{d}}{DENIS photometry (M. H. Vuong, private communication).}
\end{table*}

We compute the photometry of the {\it integrated light} of the FS using an aperture 
identical for the three filters, and background estimate from an area free of sources 
close to the disk. To avoid contamination by the close background star, 
the aperture photometry is computed from the area where the surface brightness is greater 
than 19.3\,mag\,arcsec$^{-2}$ in the three filters 
(see the boundary of the aperture photometry in Fig.~\ref{color}).
The match between the ISAAC filters used here and those used to establish 
the faint IR standard star system of Persson et al.\ (1998; a.k.a.\ the LCO photometric system) 
being quite good, we use the color transformations 
given by \cite*{carpenter01} for the LCO photometric system, 
to convert the disk magnitudes to the 2MASS photometric system. 
The resulting $J$, $H$, and $K_\mathrm{S}$ integrated magnitudes are listed 
in Table~\ref{photometry}.
These values are consistent with the values given for 2MASSI\,1628137-243139 
in \cite*{cutri00}, taking into account the photometry uncertainties. 
This implies no dramatic change in the brightness of this disk on $\sim$3\,years timescale. 
However this does not exclude very large changes over a few days like those 
seen in HH\,30\,IRS (\citealt{wood02}).

We list also in Table~\ref{photometry} existing complementary integrated magnitudes 
from other broad-band optical/NIR filters. Using the absolute calibration given by \cite*{cox00}, 
we obtain the corresponding spectral energy distribution (SED) from optical to NIR presented 
in Fig.~\ref{sed}.
This disk was not detected in the mid-infrared (MIR) by the ISOCAM survey 
although located in low-noise region of this survey
(\citealt{bontemps01}; 6 and 10.5\,mJy upper limits for 6.7 and 14.3\,$\mu$m, respectively, 
corresponding to the central wavelength of the ISOCAM filters LW2 and LW3). 
Indeed for edge-on inclinations, all the material of the disk 
midplane is on the line of sight leading to an optically thick medium in the MIR; 
continuum emission from the warm inner disk needs deeper exposure times to be detected 
(see e.g., ISOCAM MIR spectrum of HH\,30\,IRS; \citealt{stapelfeldt99}). 
The SED shape, rising in the optical and reaching a peak in $H$, is similar to that of 
HH\,30\,IRS (optical variability from \citealt{wood00}; NIR from \citealt{cotera01}; 
mid-infrared to millimetric from \citealt{brandner00} and references therein), 
and consistent with edge-on disk models (e.g. \citealt{boss96}; see also below 
\S\ref{section_sed_models}).

{
     \epsscale{1}
     \plotone{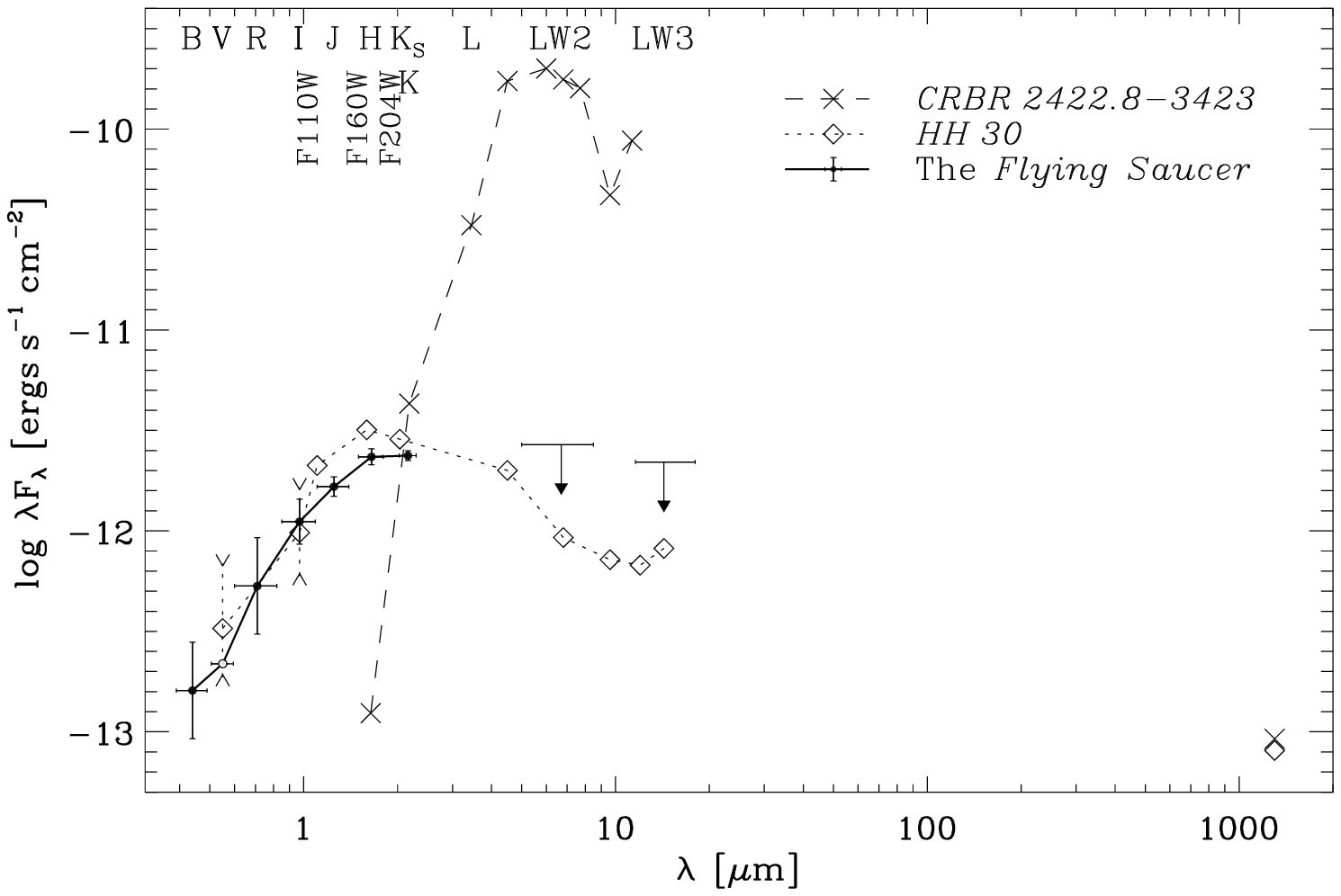}
      \figcaption{Spectral energy distributions (SED) of the Flying Saucer compared to 
others circumstellar disks. The continuous line and the dots show the SED of the Flying Saucer 
with two-sigma error bars. (The white dot is only an estimate of the $V$-band magnitude, 
see note {\it b} of Table~\ref{photometry}.)
The SED of CRBR\,2422.8-3423 located in one of the dense core of the $\rho$ 
Ophiuchi cloud (Brandner et al. 2000), and HH\,30\,IRS located in the Taurus dark 
cloud, are plotted for comparison. 
}
         \label{sed}
\vspace{0.25cm}
}

Since the FS falls outside the region mapped in the millimetric continuum by \cite*{motte98} 
we cannot have a direct measurement of the amount of circumstellar material.
An estimate of the disk mass will be given from scattered light model in \S\ref{section_model}.

\section{Foreground visual extinction}
\label{section_extinction}

     \begin{figure*}[t]
    \epsscale{2}
    \plotone{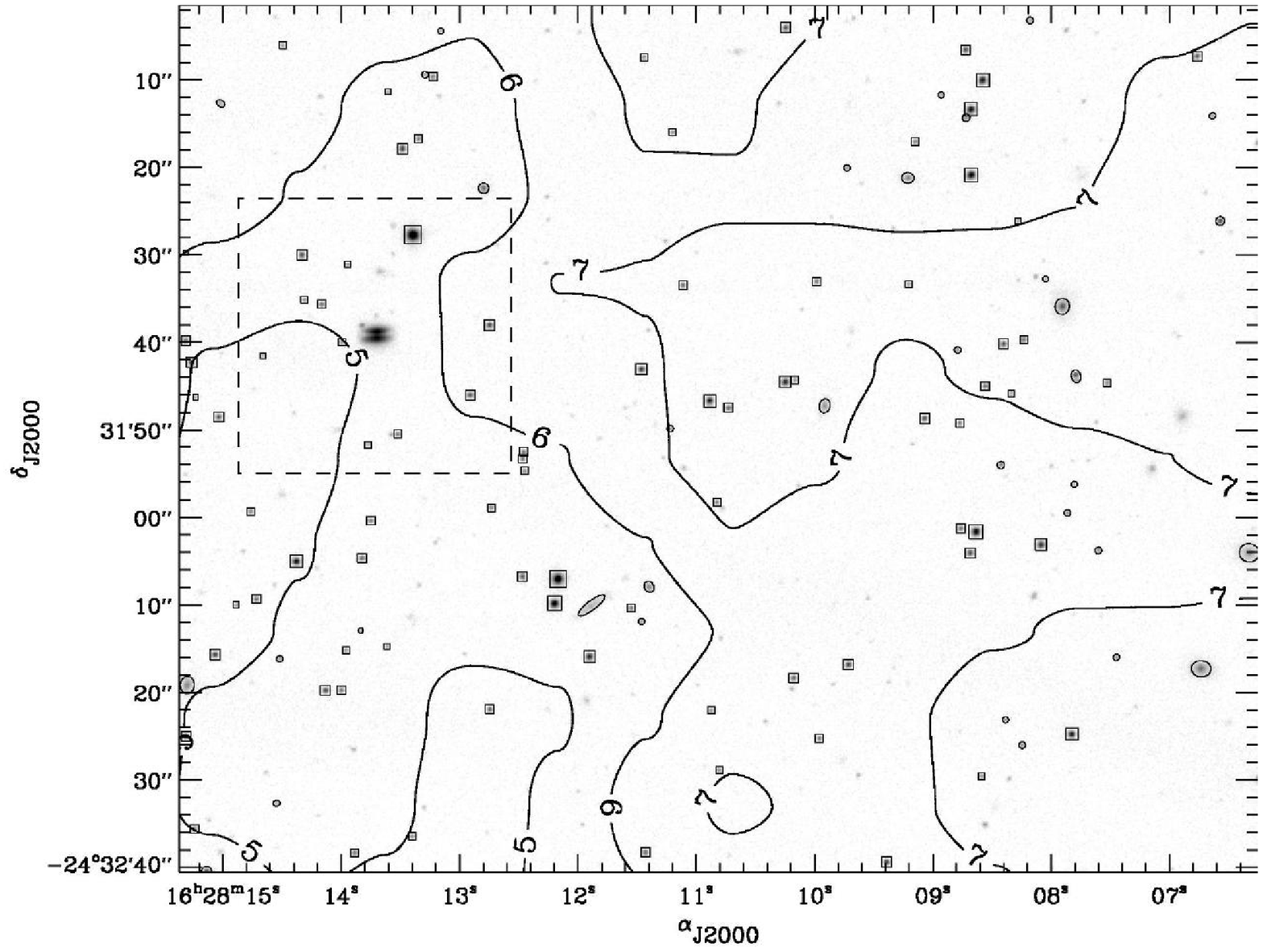}
     \caption{Extinction map of the area around the Flying Saucer. 
The background image is the $K_\mathrm{S}$-band VLT-UT1/ISAAC image 
displayed using a logarithmic color stretch. The dashed box corresponds to the area of Fig.~\ref{image}. 
Squares and ellipses mark stars and galaxies, respectively, with signal-to-noise greater 
than 5, selected using SExtractor (Bertin \& Arnouts 1996). Contours show the extinction map 
computed from the colors of the selected (background) stars. Contour labels give the visual 
extinction in magnitude. 
}
         \label{extinction_map}
\notetoeditor{Fig.~\ref{extinction_map} must be printed on more than one column with the caption on the right.}
   \end{figure*}

To estimate the foreground visual extinction of our object, $A_\mathrm{V}$, 
we have to deredden its integrated NIR colors, derived from the integrated 
magnitudes established in the previous section. 
\cite*{meyer97} observed in Taurus-Auriga that {\it dereddened} Classical T~Tauri stars (CTTS)
follow a narrow segment locus in a $J-H, H-K$ color-color diagram. 
We adopt this locus as statistically representative of the CTTS colors. 
We stress that an estimate of $A_\mathrm{V}$ based on these typical CTTS colors 
is however only indicative because the NIR light of the FS is almost 
purely scattered. Its color might then be quite different 
from that of a typical CTTS in which one primarily sees direct 
light from the stellar photosphere and the inner disk.
The color transformation of the CTTS color locus  
from the CIT to the 2MASS photometric system using \cite*{carpenter01} 
leads to: 
\begin{equation} 
(J-H)_0=0.61\pm0.12 \, (H-K_\mathrm{S})_0+0.50\pm0.07\,. 
\label{eq:locus}
\end{equation}
\cite*{meyer97} used the extinction law from \cite*{cohen81}, without 
taking into account its uncertainties, to deredden their CTTS sample. 
To be consistent with their results we take the same extinction law 
transformed to the 2MASS photometric system:
\begin{equation} 
(J-H)-(J-H)_0=0.118\pm0.001 \, A_\mathrm{V},
\end{equation}
\begin{equation} 
(H-K_\mathrm{S})-(H-K_\mathrm{S})_0=0.067\pm0.001 \, A_\mathrm{V}, 
\label{eq:extinction_law}
\end{equation}
which leads to $E(J-H)/E(H-K_\mathrm{S})=1.77\pm0.04$.
Straightforward algebra with formulae \ref{eq:locus}--\ref{eq:extinction_law} 
using the observed NIR colors ($J-H= 1.15\pm0.08$\,mag, $H-K_\mathrm{S}=0.80\pm0.06$\,mag), 
and combining the quoted uncertainties, gives $A_\mathrm{V}$=2.1$\pm$2.6\,mag 
(see Fig.~\ref{color}).

In order to obtain an upper limit on the foreground visual extinction of the FS, 
we computed the extinction map of the area around it by applying 
the ``Near Infrared Color-Excess'' (NICE) method (\citealt{lada94}; 
\citealt{alves98}). We extracted with SExtractor (\citealt{bertin96}) sources with 
signal-to-noise greater than 5 in the three filters. 
To compute this map we selected only sources with stellarity indexes greater than 0.95
(other sources were considered as galaxies).
Because there were no observations of the unreddened
background stellar field (necessary for the determination of the zero
point of the extinction scale), we must adopt a $<H-K_\mathrm{S}>$ color for
this background. \cite{alves98} found
$<H-K_\mathrm{S}>\, = 0.20\pm0.13$\,mag towards a complex region
in Cygnus, essentially at the Galactic plane, while \cite*{alves01} 
found $<H-K_\mathrm{S}>\, = 0.12\pm0.08$\,mag towards a clean
line-of-sight towards the Galactic bulge, at $b \simeq 7^{\circ}$. The
stellar background towards the Ophiuchus complex is probably even
better behaved, because it lies at $b \simeq 15^{\circ}$. On the other
hand, the $<H-K_\mathrm{S}>$ color of the North Galactic pole from 2MASS
data is also $\sim $0.12\,mag (M. Lombardi, private communication), so
we decided to adopt this as the mean Ophiuchus background color.

We show our extinction map in Fig.~\ref{extinction_map}. 
The spatial resolution of the map is 20\arcsec, and the visual extinction 
ranges from 5 to 7\,mag. 
The extinction increases from East to West, i.e.\ towards the molecular 
core (see Fig.~\ref{chart}), hence tracing the presence of material related 
with the main cloud. 
This extinction map leads to the following upper limit on the foreground 
extinction of the FS: $A_\mathrm{V} \le 5.1\pm1.2$\,mag. 
Our previous estimate based on the CTTS colors, $A_\mathrm{V}$=2.1$\pm$2.6\,mag, 
is consistent with this upper limit, and might imply that the FS is located within 
this peripherical material, which is not unreasonable. We adopt this extinction value 
as our best guess.

Thanks to its location at the periphery of the dense cores, 
this disk suffers less foreground extinction than the other disks already found 
in this star-forming region. 
This is also visible in Fig.~\ref{sed} where its SED shows 
only a slow rise at short wavelength in contrast to the steep rise 
of CRBR\,2422.8-3423 due to the high extinction of dense cores (\citealt{brandner00}).
\cite*{cotera01} found a comparable foreground visual extinction for HH\,30\,IRS 
($A_\mathrm{V} \le 4.1\pm0.9$\,mag) assuming a typical stellar type M0 
for the central star.  
Since the FS suffers no background emission from the dense cores, 
it is an ideal laboratory for the study of both dust and gas in circumstellar disks.

\section{Pixel NIR Color Mapping\\
of the reflection nebulae}
\label{section_color}

    \begin{figure*}[p]
   \plotone{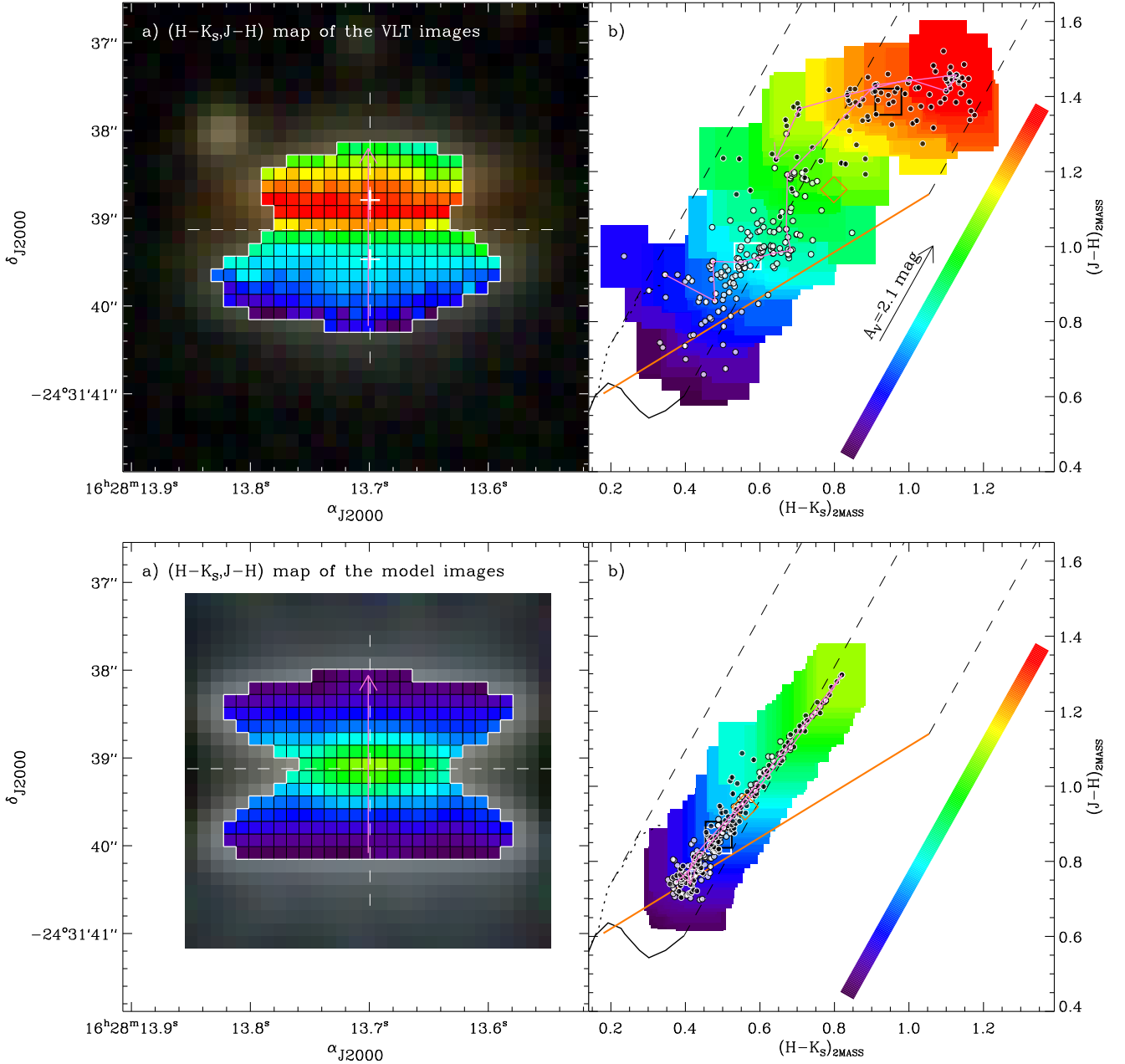}
      \caption{
Pixel NIR Color Mapping.
{\it Upper panels:} 
a) ($H-K_{\rm S}$,$J-H$) map of the VLT images.
The background image is an enlargement of the color composite image presented in 
Fig.~\ref{image}.b, resized to the field of view of Fig.~\ref{brightness}. 
The white crosses mark the peak positions 
of the reflection nebulae in the $K_\mathrm{S}$-band image, 
and the dashed lines are the symmetry axes drawn in Fig.~\ref{brightness}.
The white contour defines the boundary of the aperture photometry
defined by surface brightness at least greater than 19.3\,mag\,arcsec$^{-2}$ 
in the three NIR filters.
The color of each pixel of this area is coded according to its  
position in the $J-H, H-K_\mathrm{S}$ diagram on the right-hand panel.
The purple arrow shows the path when one moves on the vertical axis of the disk; 
the corresponding path is also shown in the right-hand panel.
b) Corresponding $J-H, H-K_\mathrm{S}$ diagram of the nebula pixels (2MASS photometric system). 
The black and dotted continuous lines show the intrinsic colors of A0--M6 dwarfs 
and giants (Bessel \& Brett 1988), respectively. 
The orange line is the CTTS locus from Meyer et al. (1997). 
Reddening vectors (Cohen et al. 1981), adapted for the 2MASS photometric system, 
are drawn for giants, M6 stars, and CTTS. 
The orange diamond shows the NIR colors of the integrated light of the disk.
The arrow marks the estimate of the disk foreground visual extinction obtained from the 
CTTS locus (see \S\ref{section_extinction}), and gives the extinction scale. 
The white and black squares mark the NIR colors of the integrated light of the southern 
and northern nebula, respectively. 
White and black dots mark the NIR colors of pixels of the southern and northern nebula, respectively.
The one-sigma photometric errors are given by the sizes of the error boxes centered on the dots.
The color palette defined on the right parameterizes the NIR colors of the nebula pixels 
in terms of simple ISM extinction.
The color of the error boxes are coded according to their projections on the color palette. 
The purple arrow corresponds to the NIR color variations when one moves on the vertical axis 
of the disk. See detailed discussion of this Fig. in \S\ref{section_color}.
{\it Lower panels:} 
a) ($H-K_\mathrm{S}$,$J-H$) map of the disk plus diffuse infalling envelope model. 
b) Corresponding $J-H, H-K_\mathrm{S}$ diagram of the model image pixels. 
Symbols are the same than used in the upper panels. 
The color palette is identical to the one use in the upper panel, allowing easier comparisons between 
the model predictions and the observations.}
         \label{color}
\notetoeditor{Fig.~\ref{color} must be printed in color on two column. 
Please do not decrease the figure size, for clarity the figure width 
must be equal to the textwidth.
Could you please convert the files to CMYK?}
\end{figure*}

We noted in the previous section the reddish appearance of the northern
reflexion nebula of the FS, appearing in Fig.~\ref{image}.  To our knowledge 
such a NIR color difference between the reflexion nebulae of an edge-on disk 
has never been observed.  NIR adaptive optics images of HK~Tau/c display only a
``slight change in the nebula's appearance between $J$ and $K$'' (\citealt{stapelfeldt98}); 
speckle imaging shows that the flux ratio between the two nebulae increases from 5 
at $J$ to 8 at $K$ (\citealt{koresko98}).  
To do a spatially resolved quantitative study of the NIR color differences 
between the two nebulae of the FS, we introduce a new NIR data visualization 
method called ``Pixel NIR color mapping'', PICMap for short. 
Although in this paper we apply it to the NIR photometry of the FS, this method 
is general and could conceivably be applied to other photometric bands and adapted 
to other spatially resolved objects.

The central idea is the following. We want to visualize directly in a 
single VLT image the $J-H$ and $H-K_\mathrm{S}$ colors of each pixel of the FS. 
We start by selecting the pixels of the FS located inside the boundary of the aperture 
photometry defined in \S\ref{section_photometry} (see the upper left-panel of Fig.~\ref{color}), 
and by computing their $J-H$ and $H-K_\mathrm{S}$ colors from 
the $J, H$ and $K_\mathrm{S}$ brightness maps of Fig.~\ref{brightness}. 
Then we plot this set of $J-H$ and $H-K_\mathrm{S}$ colors in a $J-H, H-K_\mathrm{S}$ 
color-color diagram, much in the way the NIR colors of individual stars 
in an association would be plotted. By analogy, we also plot in this diagram the standard 
ISM reddening vector.
The upper right-hand panel of Fig.~\ref{color} illustrates the main result: 
most of the pixels in the $J-H, H-K_\mathrm{S}$ diagram are found to be spread roughly along
the reddening vector. Therefore, we use its direction to build an
arbitrary color palette, shown to the right of the diagram. 
The last step of the method is then to create a mapping, 
both on the $J-H, H-K_S$ diagram and on the VLT image, by ``coding'' the color of each FS pixel 
according to its projection on this color palette. 
In other words, PICMap defines a correspondence between the  $J-H, H-K_S$ plane and the VLT image. 
As we shall see, this visualization method is powerful to detect any unusual trend in NIR colors 
within the FS image.

Physically, we could interpret the pixel color palette used in the PICMap as a visualization 
of the NIR colors of the nebula pixels in terms of pure extinction by interstellar medium 
type dust grains. We caution however that the grains in the disk may be grayer than interstellar 
medium type dust grains (see \S\ref{dust_grain}), and the true extinction along a line-of-sight 
through the nebulae is then likely considerably 
higher than the standard extinction derived from this simple ISM extinction law. 
In reality, the NIR colors of the FS result from a complex combination of extinction and 
scattering by disk grains, which can only be reproduced by modelling the radiative transfer 
of the stellar light through the dust, as discussed in the next section.
However the PICMap can be used to quantify the relative difference of NIR colors between 
the two reflection nebulae in term of standard ISM extinction.

In the upper panels of Fig.~\ref{color} when one moves in the VLT image 
from the edge of the southern nebula to the dark lane along the vertical axis of the disk, 
the corresponding NIR colors in the $J-H, H-K_\mathrm{S}$ diagram (white dots) follow 
roughly the direction of the ISM 
reddening vector, with an increase of $\sim$4\,mag of standard extinction. 
The same amount of standard extinction is also visible for the NIR colors of the pixels 
located between the edge of the northern nebula and the dark lane (black dots). 
However the northern nebula suffers 
an extra standard extinction of $\sim$3\,mag compared to the southern nebula.
In the $J-H, H-K_\mathrm{S}$ diagram, it is also clear that the NIR colors are at a peak 
for many pixels of the northern nebula, and display what would be called a 
``NIR color excess'' in the case of young stars.
The PICMap method allows to locate them easily spatially on the VLT image.  
They are not distributed at random in the northern nebula, but form a ridge, 0\farcs3 (40\,AU at 140\,pc) 
to the north of the dark lane and parallel to it. There is no similar feature in the southern nebula. 

In spite of the fact that the FS disk is seen nearly edge-on, it is clearly asymmetric with respect 
to the dark lane, the northern nebula being significantly redder than the southern nebula.
In the next section, we will use axisymmetric disk models with
radiative transfer, to look for a model parameter set that reproduces the observed 
NIR color asymmetry. We will apply our PICMap method to the scattered light model, 
to visualize the results and compare it with our observations.

\section{Radiative transfer models}
\label{section_model}

We adopt a axisymmetric density structure 
for the circumstellar material (i.e. symmetric with the disk midplane), 
with dust grains that reproduce the images and spectral energy distribution (SED) 
of HH30~IRS, and perform the radiative transfer simulation using Monte Carlo techniques 
(e.g., \citealt{whitney92}; \citealt{code95}) 
to compare scattered light models with the high resolution VLT images.  
For calculating synthetic SED, we use the Monte Carlo radiative 
equilibrium technique of \cite*{bjorkman01} 
as adapted for T~Tauri disks (Wood et al. 2002a, b).  

\subsection{Density structure of the circumstellar material}

	\subsubsection{Disk model}

As with previous models of disk SEDs and scattered light images, 
we adopt a axisymmetric flared disk structure 
in which the total density (gas plus dust) is:
\begin{equation}
\rho_\mathrm{disk}=\rho_0 \left({R_\star\over{\varpi}}\right)^{\alpha}
\exp{\left(-\frac{z^2}{2\,h^2(\varpi)}\right)}  
\; ,
\end{equation}
where $\rho_0$ is the disk density extrapolated to the stellar surface 
(we set $R_\star=1.03$\,R$_\odot$; see SED model \S\ref{section_sed_models}), 
$\varpi$ is the radial coordinate in the disk midplane (with $\varpi \ge R_\star$), 
and the scale height increases with radius, 
$h\mathrm{(}\varpi \mathrm{)}=h_0\left({\varpi /{R_\star}}\right)^\beta$.  
We assume that the gas and dust are well mixed throughout the disk.
We adopt for the flaring parameter $\beta$ the value derived in the hydrostatic disk 
structure models of \cite*{dalessio98}, $\beta=5/4$; and set $\alpha=3(\beta-0.5)=9/4$, 
which is the value appropriate for viscous accretion theory (\citealt{shakura73}).
This leads to a surface density $\Sigma\propto \varpi^{-1}$ (\citealt{dalessio98}), 
flatter than the one usually found in the literature ($\Sigma\propto \varpi^{-3/2}$; 
e.g., \citealt{beckwith90}).
The disk outer radius, $\varpi_\mathrm{out}$, is fixed to 300\,AU to match 
the extent found in the VLT $J$ image. 
This value is a lower limit of the disk outer radius.
In our model the disk is passive, without any source of internal heating.
For low mass disks ($M_\mathrm{disk}<<0.1$\,M$_\odot$), 
the heating due to accretion luminosity is negligible and stellar 
irradiation dominates the disk heating (\citealt{dalessio98}).  
The light coming out of this system is thus only the light of the star reprocessed by 
the disk dust grains.

	\subsubsection{Diffuse infalling envelope model}

It is necessary to add another circumstellar component because a model featuring only by 
a disk produces reflection nebulae with a flat hourglass shape, without the rounded tips visible 
in Fig.~\ref{brightness}. Inspired by the modelling of HH\,30\,IRS (\citealt{wood98}),
we have included a diffuse infalling envelope in our model.
The envelope structure follows the rotational collapse geometry of \cite*{terebey84}
in which the total density (gas plus dust) is:%
\begin{eqnarray} 
\rho_\mathrm{env} & = & \frac{\dot M_\mathrm{env}}{4\pi} 
\left(GM_\star r_\mathrm{c}^3\right)^{-\frac{1}{2}} 
\left(\frac{r}{r_\mathrm{c}}\right)^{-\frac{3}{2}} 
\left(1+\frac{\mu}{\mu_0}\right)^{-\frac{1}{2}} \nonumber \\
                  &   & \times \left(\frac{\mu}{\mu_0} + 
\frac{2\mu_0^2 r_\mathrm{c}} {r}\right)^{-1} \; ,
\end{eqnarray}
where $\dot M_\mathrm{env}$ is the mass infall rate, $r_\mathrm{c}$ is the centrifugal radius, 
$\mu = \cos\theta$ (with $\theta$ the polar angle measured from the disk axis), 
and $\mu_0=\mu(r\rightarrow \infty)$ is determined by
\begin{equation} \mu_0^3 +
\mu_0 \left(\frac{r}{r_\mathrm{c}} - 1\right) - \mu \, \frac{r}{r_\mathrm{c}} = 0 \; .
\end{equation}
For our models we adopt $r_\mathrm{c}=\varpi_\mathrm{out}=300$\,AU, $M_\star=0.5$\,M$_\odot$, 
and an outer envelope radius of 1000\,AU.  We also include a curved bipolar evacuated 
cavity in the envelope.  The cavity walls are curved 
with their height scaling as $z_{\rm cav}\sim \varpi^2$ and a half 
opening angle of $20^\circ$ at the outer radius of the envelope.

\subsection{Properties of the disk dust grains}
\label{dust_grain}

In young stellar objects, the wavelength dependence of the dust properties are 
inferred from the slope of the sub-millimetric continuum emission 
(e.g., \citealt{beckwith90}; \citealt{beckwith91}) and also from the wavelength dependence 
of the width of dust lanes of edge-on disks (e.g., \citealt{cotera01}). 
These studies show that opacity laws shallower than the one of the ISM are needed.
Indeed for HH~30~IRS, ISM type grains produce very compact images in $K_\mathrm{S}$ 
(\citealt{wood98}), without dark lanes. We use the dust grains described 
in \cite*{wood02}, which reproduce the multi-wavelength images and continuum fluxes 
for the HH\,30\,IRS disk. This dust model uses a ``power law with exponential cutoff'' 
for the dust size distribution. The exponential scale length for this dust distribution, 
$a_\mathrm{c}=50$\,$\mu$m, indicates dust grains have grown to larger than $50$\,$\mu$m 
within the HH\,30\,IRS disk. 

    \begin{figure*}[t]
    \plotone{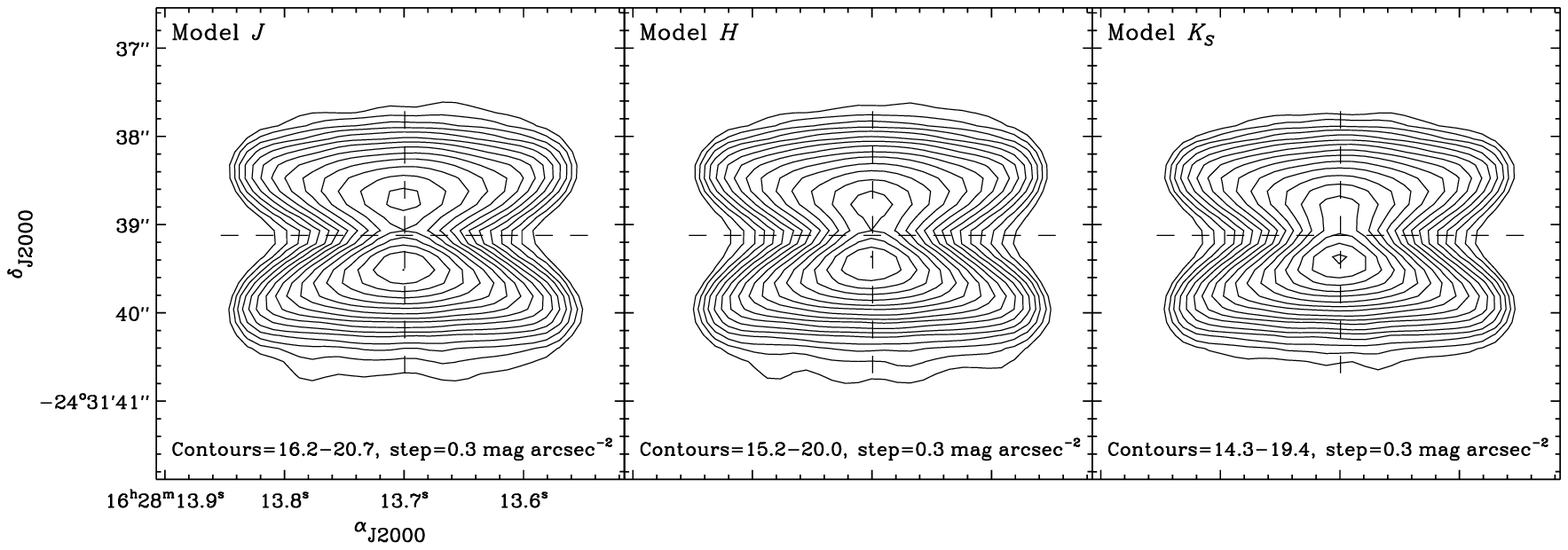}
     \caption{Surface brightness maps of our disk model (see text for the parameter values).
}
         \label{model_surface_brightness}
\notetoeditor{Fig.~\ref{model_surface_brightness} must be printed on two columns.}
   \end{figure*}

The parameters for the gas plus dust mixture for the NIR scattered light models are 
the opacity, $\kappa$, albedo, $\omega$, and Henyey-Greenstein scattering asymmetry 
parameter, $g$ (\citealt{henyey41}), and are shown in Table~2.  
The opacity for this model varies from 
$\kappa_J = 24$\,cm$^2$\,g$^{-1}$ to $\kappa_K = 16$\,cm$^2$\,g$^{-1}$, 
while for the ISM the opacity is larger and has a steeper wavelength dependence with 
$\kappa_J = 65$\,cm$^2$\,g$^{-1}$ and  
$\kappa_K = 22$\,cm$^2$\,g$^{-1}$ (e.g., \citealt{kim94}).

{
\begin{center}
Table 2\\
{\sc Parameters adopted}\\
{\sc for the gas plus dust mixture.}\\
\smallskip
\begin{tabular}{ccccccc}
\hline
\hline
\noalign{\smallskip}
Filter & $\kappa^{\tablenotemark{a}}$  & $\omega^{\tablenotemark{b}}$ & $g^{\tablenotemark{c}}$\\
band & [cm$^2$\,g$^{-1}$] \\
\noalign{\smallskip}
\hline
\noalign{\smallskip}
$J$  & 24 &  0.53 & 0.59  \\
$H$  & 19 &  0.53 & 0.58  \\
$K$  & 16 &  0.54 & 0.57  \\
\noalign{\smallskip}
\hline
\end{tabular}
\tablenotetext{\mathrm{a}}{The opacity.}
\tablenotetext{\mathrm{b}}{The albedo.}
\tablenotetext{\mathrm{c}}{The Henyey-Greenstein scattering asymmetry parameter.}
\end{center}
\vspace{0.25cm}
}

\subsection{Scattered light model of the FS}

Having set up the circumstellar material density structure and dust grain properties 
as used for modelling other circumstellar disks, we compute a synthetic scattered light image 
at $\lambda$=1.25\,$\mu$m, convolved with a 0\farcs4-FWHM Gaussian point spread function.
We do not include additional noise. 
We vary the disk inclination, $i$, the disk scale height, $h_0$, disk mass, $M_\mathrm{disk}$, 
and envelope accretion rate $\dot M_\mathrm{env}$ until we reproduce the VLT $J$ image.
Then we make scattered light models for the central wavelengths 
of the $H$ and $K_\mathrm{S}$ filter. 
The intrinsic luminosity and spectrum of the central star is not set in this calculation. 
Instead the model surface brightness is normalized in each filter using the peak surface brightness
of the southern nebula.
Among all the free parameters of our model we note that only the disk inclination 
could produce an asymmetry between the two reflection nebulae in the resulting scattered 
light images.

Figure~\ref{model_surface_brightness} shows our simulation for a 300\,AU outer radius  
disk viewed at $i=86^\circ$ with $h_0=0.015$\,R$_\star$, giving $h$(100\,AU)=15.3\,AU, 
and $M_\mathrm{disk} = 2\,10^{-3}$\,M$_\odot$. 
For comparison \cite*{wood02} find using the same modelling 
a similar disk mass for HH\,30\,IRS with $M_\mathrm{disk} = 1.5\,10^{-3}$\,M$_\odot$.
From the various models computed with the gas plus dust mixture parameters as given 
in Table~2 we estimate the following uncertainties: 
$\sigma_\mathrm{i}=1^\circ$, $\sigma_\mathrm{h(100\,AU)}=1$\,AU, 
$\sigma_\mathrm{M_\mathrm{disk}} = 0.5\,10^{-3}$\,M$_\odot$.
The disk mass we derive is obviously sensitive to the dust grain properties. 
For instance for fixed values of $\omega$ and $g$, what is really determined from 
the modelling is the product of disk mass and opacity. 
If we were to use a different dust opacity the disk mass would scale according to that.
Moreover the disk mass is a lower limit since the opacity is not sensitive 
to very large particles such as rocks, planetesimals, or planets. 
The envelope is very tenuous, with a total mass $M_\mathrm{env}=4\times 10^{-4}$\,M$_\odot$.  
The envelope accretion rate is very weak with 
$\dot M_\mathrm{env} = 10^{-7}$\,M$_\odot$\,yr$^{-1}$.  
For comparison typical envelope accretion rates of evolved protostars, 
Class~I sources (\citealt{lada91}), 
are larger with $\dot M_\mathrm{env} = (2$--$10) \times 10^{-6}$\,M$_\odot$\,yr$^{-1}$ 
(\citealt{strom94}; \citealt{kenyon93}; \citealt{whitney97}), 
whereas CTTS have  $\dot M_\mathrm{disk} = 10^{-8}$--$10^{-6}$\,M$_\odot$\,yr$^{-1}$ 
(\citealt{strom94}, and references therein).
The circumstellar material of the FS is thus characteristic of a CTTS.
Our symmetric disk model reproduces correctly the North-South extent of the FS 
in the three bands thanks to this diffuse infalling envelope, and the East-West shape 
of the southern nebula. 
However the width of the northern nebula appears $\sim25\%$ larger than 
in the VLT data. A smaller inclination ($\sim80^\circ$) improves the match with 
the observation by reducing the width of the northern nebula, but decreases 
also dramatically its peak brightness leading to a more important discrepancy. 

There are also discrepancies between our axisymmetric model and the VLT observation 
as we go to longer wavelengths. In particular the width of the dark lane 
is reproduced only in the $J$-band. At longer wavelengths the width of the dark lane 
in the axisymmetric models decreases as the dust opacity decreases, and the brightness of 
the northern nebula is always less than that of the southern nebula (e.g., \citealt{wood98}).  
However, the VLT data show no variation of the width of the dust lane with the wavelength, 
and moreover as already noted in \S\ref{section_photometry} 
the northern nebula brightness relative to the southern nebula brightness {\it increases}.
A steeper grain size distribution and/or smaller maximum grain size may explain the absence 
of variation of the width of the dark lane at NIR (\citealt{dalessio01}), but cannot 
reproduce the brightening of the northern nebula.
The determination of an adapted dust opacity law for this object is beyond the scope of 
this paper, and should be inferred using complementary observation made at longer 
wavelengths, more constraining for the dust size distribution (e.g., \citealt{wood02}).
 
The lower panels of Fig.~\ref{color} shows the resulting PICMap for our model.
We can see easily with this tool that the NIR color distribution of the southern nebula 
along the reddening vector is quite well reproduced, however the NIR color distribution of 
the two nebulae is nearly symmetric with the dark lane. 
We conclude that the {\sl axisymmetric} disk model does not reproduce the asymmetry 
in the color spatial distribution observed in the scattered light of the FS.

\subsection{SED model of the FS}
\label{section_sed_models}

Since the spectral type of the central star is unknown, 
we have assumed $T_\mathrm{eff}=3500$\,K, and used the corresponding 
Kurucz atmosphere model (\citealt{kurucz94}) for the input stellar spectrum.  
As the envelope has a small mass compared to the disk mass, distributed throughout 
a much larger volume than the denser disk, we neglect it and we compute the synthetic 
emergent SED from our symmetric disk model only.
We apply the foreground visual extinction estimated in \S\ref{section_extinction} 
using the analytical fits from optical to IR of the extinction cross-section given by 
\cite*{ryter96}, combined with the canonical relation between visual extinction and column 
density obtained by \cite*{predehl95}.
Adopting a source distance $d=140$\,pc we require that $L_\star=0.14$\,L$_\odot$ 
(which corresponds to $R_\star= 1.03$\,R$_\odot$) to reproduce both the optical 
and the NIR fluxes (Fig.~8).
Below 2\,$\mu$ the SED is dominated by the (scattered) light of star, while 
at longer wavelengths the emission from the disk dust grains dominates.
Varying the disk inclination from pole-on to edge-on reduces the flux 
because the disk midplane is optically thick in the MIR.
Our SED simulation predicts 10--20\,$\mu$m fluxes of 0.7--1.8\,mJy, and 1.3\,mm 
fluxes of 12.5\,mJy. 
\newpage

{
\vspace{0.25cm}
	\epsscale{1}
      \plotone{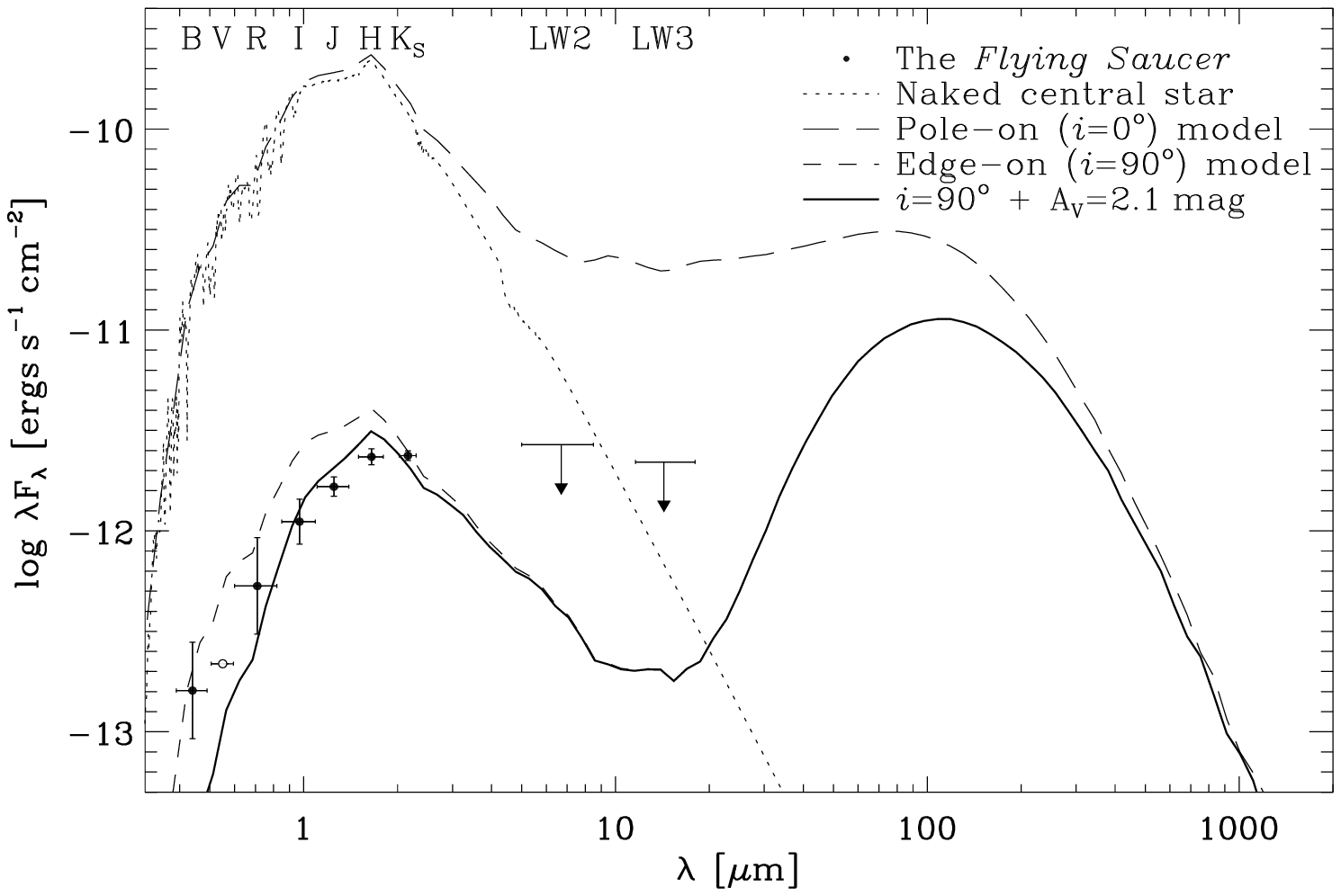}
      \figcaption{Spectral energy distribution for the Flying Saucer model.
The black dots mark the observed optical and NIR fluxes with two-sigma error bars. 
(The white dot is only an estimate of the $V$-band magnitude, 
see note {\it b} of Table~\ref{photometry}.)
The continuous line shows our best model: an edge-on disk (short-dashed line) 
plus 2.1\,mag foreground visual extinction. 
The dotted line is the Kurucz atmosphere model for $T_\mathrm{eff}=3500$\,K 
taken as the spectrum of the central star. 
The long-dashed line is the emergent SED when the disk is viewed pole-on.  
}
\vspace{0.25cm}
}

\section{Discussion}
\label{section_discussion}

The spatial study of circumstellar disk is usually based only on surface brightness 
maps as in Fig.~\ref{brightness}. Thanks to the high spatial resolution achieved with 
our VLT NIR observations (0\farcs4-seeing images), we have been able to make a spatial 
study of the NIR colors of the individual pixels of the two resolved reflection nebulae 
of this nearly edge-on disk. 
We introduced in this paper a new NIR data visualization called ``Pixel NIR color mapping'' 
(PICMap for short), to visualize directly the NIR colors of the nebula pixels. 
This is a power tool to identify and localize any unusual NIR color trend in the reflection nebulae.
We have found with it a NIR color excess and an extra standard extinction 
in the northern reflection nebula.
Fitting the images and the SED with axisymmetric disk model leads to model 
parameters consistent with what would be expected for a typical CTTS, but cannot reproduce 
the observed NIR colors. It is necessary to introduce a source of asymmetry 
in our axisymmetric modelling.

We use in our model a uniformly bright star. There could be cool or dark spots 
on the star that will change the illumination pattern and consequently 
the brightening of the northern nebula. Stellar hotspot models 
produce a typical lighthouse effect, which has been proposed to explain 
the observed brightening and dimming of the reflection nebulae 
in HH\,30\,IRS (\citealt{wood_whitney98}).
However such a phenomenon should produce changes at least on a stellar rotation period 
($\sim$few days), it is thus unlikely to be observed at the same phase on a 
130\,day time interval. We could have an embedded shock associated 
with a micro-jet, which could emit H$_2$ lines producing the 
brightening in $H$ and $K_\mathrm{S}$ (e.g., \citealt{grosso01}). However there is no evidence 
in this object of H$_2$ emission knots outside the nebulae. This micro-jet hypothesis 
in the northern nebula can be easily tested with NIR spectroscopy and narrow band filter imaging. 

We could account for the lower brightness of the northern nebula at short wavelengths 
and its redder infrared color as due to $\sim3$\,mag of localized foreground extinction.
From the color-color diagram of the model image (see lower right-hand panel of Fig.~\ref{color}), 
we note that adding an extra extinction of $\sim3$\,mag only in front of 
the northern nebula reproduces the observed range of $J-H$ colors. 
However this localized foreground extinction is not visible in our extinction map 
(Fig.~\ref{extinction_map}). Adding $\sim3$\,mag foreground extinction only above the dark lane 
can only be an ad hoc solution. 
A more natural way to explain these differences would be to 
introduce an extra self-absorption, possibly due to a large-scale disk warp. 
For instance \cite*{launhardt01} have detected the millimeter range the presence of 
a symmetric 20$^\circ$ warp beyond 120\,AU in an $\sim200$\,AU-radius edge-on disk. 
If a such large-scale disk warp is seen in front of the central star, it could strongly 
affect the scattered light.
Scattered light models of warped disks are now needed to obtain quantitative 
predictions. This idea would be fruitful to test in detail, as a large-scale disk warp 
may be excited by a planet on an inclined orbit (e.g., \citealt{lubow01}). 

We stress that our axisymmetric model assumes that the dust and gas are well mixed, 
it does not take into account the settling of the larger grains in the disk midplane. 
Moreover, having no observational constraints at longer wavelengths, we use conservatively 
a gas plus dust mixture that reproduces the images and SED of HH30~IRS. 
Other grain properties combined with dust settling might possibly account for 
some of the observed color gradient.

In summary, while we can reproduce many of the features of the FS using a symmetric 
disk plus envelope, our models should be extended to include non-symmetric geometries, 
specific dust properties, and dust settling, 
to explain the shortcomings of the model we have presented.  
The PICMap method introduced in this paper is a power tool
to compare the predictions of these improved models with the strong spatial constraints brought 
by our observations.
A more extensive coverage of the SED of the FS from NIR to MIR and millimetric, combining ground 
and space observations (e.g., HST, VLT/NACO-LGS, SIRTF, Plateau de Bure Interferometer),
 will further help to constrain the dust properties (e.g., \citealt{dalessio01}; 
\citealt{wood02}) and the disk structure of this object.

\begin{acknowledgements}
We would like to thank the referee, Chris Koresko, 
for having delivered rapidly his report, 
and for suggestions and comments that improved the final manuscript.
We also thank the Director of ESO, Catherine Cesarsky, 
who gave us a part of her discretionary time to carry out this work, 
the anonymous observer who triggered the VLT service mode observation,
and Anne Dutrey, Fran{\c c}ois M{\'e}nard, and Barbara Whitney for useful discussions.
We acknowledge financial support from the European Union (HPMF-CT-1999-00228; N.~G.);  
the PPARC Advanced Fellowship (K.~W.);  
the BMBF (DLR grant 50 OR 0003; R.~N.); 
the NASA's Long Term Space Astrophysics Research Program (NAG5~3248; J.~E.~B.), 
and the National Science Foundation (AST~9819928; J.~E.~B.).
This publication makes use of data products from the Two Micron All Sky Survey, 
which is a joint project of the University of Massachusetts and
the Infrared Processing and Analysis Center/California Institute of Technology, 
funded by the NASA and the National Science Foundation. 
We also used the SuperCOSMOS Sky Survey of the Institute for Astronomy, Royal Observatory, 
Edinburgh.
\end{acknowledgements}

\end{document}